\documentclass[preprint2,longabstract]{aastex}

\usepackage{url}

\usepackage{color}
\usepackage{verbatim}

\newcommand{\mone}  {^{-1}}
\newcommand{\mtwo}  {^{-2}}
\newcommand{\mthree}{^{-3}}

\newcommand{\cmmtwo}{\,\mathrm{cm\mtwo}}

\newcommand{\cmmthree}{\,\mathrm{cm\mthree}}

\newcommand{\eflux} {\,\mathrm{erg\,cm\mtwo\,s\mone}}

\newcommand{\kev}   {\,\mathrm{keV}}
\newcommand{\kpc}   {\,\mathrm{kpc}}

\newcommand{\kms}   {\,\mathrm{km\,s\mone}}
\newcommand{\ks}    {\,\mathrm{ks}}

\newcommand{\lum}   {\,\mathrm{erg\,s\mone}}

\newcommand{\mk}    {\,\mathrm{MK}}
\newcommand{\msun}  {\,M_\odot}

\newcommand{\chan}  {{\it Chandra}}

\newcommand{\xmm}{{\em XMM-Newton}}
\newcommand{\cxo}{{\em Chandra}}

\newcommand{\msim}{\raisebox{-.4ex}{$\stackrel{>}{\scriptstyle \sim}$}}

\newcommand{\myr}{\mbox{$M_\odot\,\mathrm{yr}^{-1}$}}

\newcommand{\cyg}{{Cyg\,OB2\,12}}
\newcommand{\hip}{HIP~101364}
\newcommand{\vicyg}{VI~Cyg~12}

\newcounter{ion}
\newcommand{\eli}[2]{\setcounter{ion}{#2}#1{\,\sc\roman{ion}}}

\def\changed{}
\def\nchange{}

\slugcomment{The scientific results reported in this article are based  on 
observations made by the Chandra X-ray Observatory.}

\shorttitle{X-ray observations of \hip}
\shortauthors{Oskinova et al.}

\begin{document}


\title{On the binary nature of massive blue hypergiants: high-resolution 
X-ray spectroscopy suggests that \cyg\ is a colliding wind binary}

\author{L.M. Oskinova}
\affil{Institute for Physics and Astronomy, University Potsdam, 
14476 Potsdam, Germany\\
\email{lida@astro.physik.uni-potsdam.de}}
\author{D.P. Huenemoerder}
\affil{Massachusetts Institute of Technology, Kavli Institute 
for Astrophysics and Space Research, 70 Vassar St., Cambridge, 
MA 02139, USA}
\author{W.-R. Hamann} 
\affil{Institute for Physics and Astronomy, University Potsdam, 
14476 Potsdam, Germany}
\author{T. Shenar} 
\affil{Institute for Physics and Astronomy, University Potsdam, 
14476 Potsdam, Germany}
\author{A.\,A.\,C. Sander} 
\affil{Institute for Physics and Astronomy, University Potsdam, 
14476 Potsdam, Germany}
\author{R. Ignace}
\affil{Department of Physics and Astronomy, East Tennessee State 
University, Johnson City, TN 37663, USA}
\author{H. Todt} 
\affil{Institute for Physics and Astronomy, University Potsdam, 
14476 Potsdam, Germany}
\author{R. Hainich} 
\affil{Institute for Physics and Astronomy, University Potsdam, 
14476 Potsdam, Germany}

\begin{abstract}
{\changed The blue hypergiant \cyg\ (B3Ia$^+$) is a representative member of
the class of very massive stars in a poorly understood evolutionary
stage. We obtained its high-resolution X-ray spectrum using {\em Chandra} 
observatory. PoWR  model atmospheres were calculated to provide realistic wind 
opacities  and to establish the wind density 
structure. We find that collisional de-excitation is the dominant mechanism 
de-populating the metastable upper levels of the forbidden lines of the He-like 
ions Si\,{\sc xiv} and Mg\,{\sc xii}. Comparison between the model and 
observations reveals that X-ray emission is produced in a dense plasma, 
which could reside only at the photosphere or in a colliding wind zone
between binary components. The observed X-ray spectra are well fitted by 
thermal plasma models, with average temperatures in excess of 10\,MK. The wind
speed in \cyg\ is not high enough to power such high temperatures, but
the collision of two winds in a binary system can be sufficient. We
used archival data to investigate the X-ray properties of other 
blue hypergiants. In general, stars of this class are not detected as
X-rays sources. We suggest that our new \cxo\ observations of \cyg\ can
be best explained if \cyg\ is a colliding wind binary possessing a late
O-type companion. This makes \cyg\ only the second binary system among 
the 16 known Galactic hypergiants. This low binary 
fraction indicates that the blue hypergiants are likely products of massive 
binary evolution during which they either accreted a significant amount of mass 
or already merged with their companion.}  
\end{abstract}


\section{Introduction}

Only a small number of stars have established masses in excess of
$100\msun$. Albeit such heavy-weights are rare, their formation,
evolution, and deaths are of significant interest.

Very massive stars have the highest bolometric luminosities among all
stellar  types, and are often located  above the empiric
Humphreys-Davidson limit  that restricts the domain of stable stars
\citep{Hump1979} in the Hertzsprung-Russell diagram (HRD). 
Their extremely large luminosities  allow us to observe them at
large distances as well as in heavily obscured regions 
\citep{Barniske2008,Crow2010}. 

{\changed Very massive stars display a rich variety of phenomena 
(e.g.\ luminous blue  variables, LBVs) and populate various 
spectral types.  Among them are late-type 
nitrogen-sequence Wolf-Rayet stars (WNh), Of/WNL transition type stars, and the 
blue hypergiants with luminosity class Ia$^+$. The latter class is the subject 
of this study. The evolutionary relationship between various types of very 
massive stars is not yet established, despite large theoretical efforts
\citep[e.g.][]{Sanyal2015}.}

Mass loss plays a major role in determining the evolutionary path of
very massive stars. These stars may lose matter via three mechanisms.
First, by line-driven winds as is ubiquitous for all hot massive stars
\citep[e.g.][]{graf2005}. Alternatively, mass can be lost via
super-Eddington winds or LBV eruptions \citep[][]{Shaviv2000,
Quataert2016}. In binary systems, finally, mass transfer may occur that
significantly affects the evolution \citep{Vanbev1998,Langer2012}.
Which of these mechanism is most important in the lives of blue
hypergiants is not yet clear.

In the present paper we employ X-ray observations to investigate the 
nature of one of the most massive and luminous stars in the Milky Way,
\cyg\ (\object \hip, \vicyg, Schulte 12). Over the last decade, X-rays
have become an established diagnostic tool to probe massive stars
\citep[][and  references therein]{Osk2016}.

Radiatively driven winds of early-type stars are typically fast
($v_\infty  >1000$\,km\,s$^{-1}$) and intrinsically unstable
\citep{Lucy1980, feld1997}. In the winds from OB-type supergiants, 
hydrodynamical 
instabilities lead to strong shocks where some fraction of the wind 
is heated to a few million Kelvin. The X-ray luminosity of these stars 
correlates with their bolometric luminosity as $L_{\rm X}\propto
10^{-7} L_{\rm  bol}$ \citep{Pal1981}. The X-ray spectra of OB stars
are thermal. The  hottest plasma is generated in the inner wind
regions, and the  X-ray lines are broad and often blue-shifted 
\citep{osk2006,WC2007,Herve2013,Puebla2016}. The X-ray luminosity
displays slow variability on the time-scale of days and on the level of
a few times ten percent \citep{osk2001,naze2013,Ignace2013,Massa2014}. In
stars with very dense winds, such as e.g.\ Wolf-Rayet (WR) stars, X-rays may be 
produced by the interaction of the fast wind flow with slower wind 
structures far out in the wind \citep{osk2012, Gayley2016}. This
mechanism is manifested by broad and blue-shifted X-ray line profiles 
and associated plasma temperatures of $\msim 10$\,MK \citep{Hue2015}.   

\begin{figure}[t]
\begin{center}
\includegraphics[width=0.99\columnwidth]{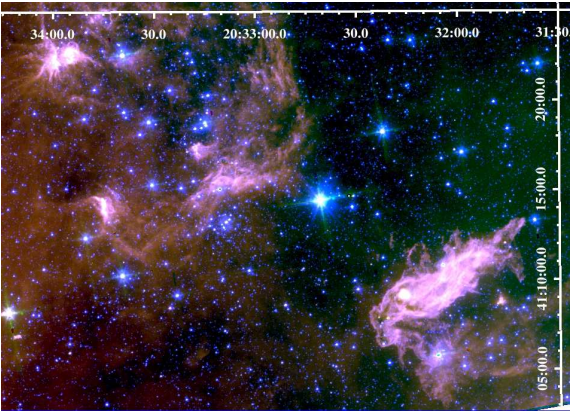}
\caption{Color composite  Spitzer IRAC image
(blue: 3.6$\mu m$, green: 4.5$\mu m$, red: 8.0 $\mu m$) with
\cyg\ (bright star close to  the center. The image size
is 24\farcm$\times$27\farcm. North is up and east is left.}
\end{center}
\label{fig:noneb}
\end{figure}

The majority of massive stars are found in binary or multiple systems 
\citep[e.g.][]{Chini2012,Sana2012}. In such systems, the winds of the
components may collide, leading to the emission of X-rays from the wind
collision zone \citep[e.g.][]{Pittard2009}. The X-ray  signatures
of colliding wind binaries are well established \citep{Rauw2016}. 
The plasma temperature is often higher than that measured from the X-ray
spectra of single stars. Colliding-wind X-ray spectra may show signs 
that the plasma departs from collisional ionization equilibrium
\citep{Pollock2005}. The X-ray line profiles may display a variety of 
shapes \citep{Henley2005}. The X-ray light curves of colliding-wind 
binaries typically show orbital variability. 

However, not all hot massive stars are detectable X-ray sources.
{\changed E.g.\, single metal enriched WR stars with spectral types WO and WC  
are quite weak in X-rays \citep{Osk2009, Rauwc2015}. Single LBVs  are
also weak X-ray sources.} The winds of these
stars are slow (a few$\times 100$\,km\,s$^{-1}$) and dense 
($\dot{M}=10^{-3}\,...\,10^{-6}$\,\myr) \citep[e.g.][]{Hillier2001}. The
radiative driving instabilities and associated shocks, likely, do not
develop in these winds. Moreover, the high wind density makes it
especially difficult for the X-rays to escape. The lack of X-ray
emission is observationally established for the majority of LBVs. All
LBVs which have been found to be X-ray bright are colliding wind
systems \citep{osk2005,naze2012}.

\begin{table*}
  \begin{center}
    \caption{Fundamental parameters of \cyg\ 
             (spectral type  B3--4 Ia$^+$)}
    \label{tab:par}
    \medskip
    \begin{tabular}{ccccccccccc}
      \hline\hline 
      \rule[-1mm]{0mm}{5mm}$T_\ast^{\rm a}$&  
      $R_\ast^{\rm a}$&
      $M_\ast^{\rm a}$& 
      $\log \dot{M}^{\rm a}$&
      $v_\infty^{\rm a}$ &
      $d$&
      $E_{{\rm B}-{\rm V}}^{\rm b}$ & 
      $\log L_\mathrm{bol}^{\rm b}$&
      $M_\mathrm{V}^{\rm b}$& 
      $\Gamma_\mathrm{Edd}$&
      $\log L_\mathrm{x}$ \\
      $[$kK$]$&
      $[R_\odot]$&
      $[M_\odot]$&
      $[M_\odot\,\mathrm{yr}^{-1}]$& 
      $[\kms]$&
      [kpc]&
      [mag]&
      $[L_\odot]$&
      [mag]&
      &
      $[\mathrm{erg\,s^{-1}}]$ \\[1mm]
      \hline 
      13.7 \rule[-1mm]{0mm}{5mm}&
      $229$&
      $110$&
      $-5.52$&
      $400$&
      1.75 &
      $3.33$&
      $6.22$& 
      $-9.82$&
      $0.38$&
      $33.8$\\
      \hline 
    \end{tabular}\\[2mm]
{\small $^{\rm a}$ stellar and wind parameters adopted from
\citet{Clark2012}; $^{\rm b}$ $E_{{\rm B}-{\rm V}}$ taken from the paper by 
\citet{Whittet2015}, which led in consequence to a slight revision of 
$L_\mathrm{bol}$ and $M_\mathrm{V}$ (see text). $L_\mathrm{x}$ 
refers to the $0.2$--$10.0\kev$ band.} 
  \end{center}
\end{table*}


In this paper we investigate the nature of the X-ray emission of \cyg\
by means of high-resolution X-ray spectroscopy. The object of our
study is introduced more fully in Sect.\,\ref{sec:hip}. The \cxo\ X-ray 
spectrum is addressed in Sect.\,\ref{sec:obs}, and the 
conclusions from our study are presented in Sect.\,\ref{sec:disc}.

\section{\cyg}\label{sec:hip}

\cyg\ is among the most massive Galactic blue hypergiants known.  This
star is a likely member of the Cyg\,OB2 association. In agreement with 
\citet{Clark2012}, we adopt a distance of $1.75\kpc$ throughout this
work. \cyg\ suffers significant reddening ($A_\mathrm{V} \approx
10\,\mathrm{mag}$). Part of this high extinction could be due to  
circumstellar matter that might have been lost
over the evolution of this very massive star.  \citet{Mar2016}
suggested that such a circumstellar shell could absorb
up to $1\,\mathrm{mag}$ in the V band. \citet{Whittet2015} pointed out 
that the properties of the interstellar matter (ISM) towards \cyg\ are not
special, while \citet{Gredel2001} suggested that the significant X-ray
luminosity of \cyg\ may affect the ISM in its vicinity.  

Nebulae are commonly observed around LBV and post-LBV stars as well as
WR stars \citep[e.g.][]{Toala2015,Steinke2016}.   {\changed  
\citet{Kobul2012} have considered  IR and mm emission from Cyg\,OB2 
region, but they do not report on a circumstellar nebular around \cyg.} 
To further search for circumstellar  matter around \cyg, we scrutinized 
archival data obtained by the {\em Spitzer} infra-red telescope; however, no
circumstellar nebula heated by the intense radiation of \cyg\ is
evident (see Fig.~\ref{fig:noneb}).

\subsection{Stellar and wind parameters}\label{sec:windpar}

\citet{Clark2012} provided a comprehensive study of \cyg. Stellar and
wind parameters were derived from the analysis of optical spectra by
means of a non-LTE stellar atmosphere model.  For the wind velocity, a
typical $\beta$-law was used, with $v(r) = v_\infty (1-b/r)^{\beta}$,
where $b\sim 1$ is a parameter that ensures a smooth 
connection between the $\beta$-law regime and the hydrostatic layer. For
the terminal wind velocity,  \citet{Clark2012} adopted $v_\infty =
400\kms$, while noting that any values between $300$ and $1000\kms$
could not be strictly ruled out. Similarly, it was noticed that while
values of $\beta$ below 2 or above 4.5 could not be excluded, the best
line fits were obtained for $\beta=3$.

For modeling the atmosphere and wind of \cyg,  we made use of the
non-LTE stellar atmosphere code PoWR
\citep[e.g.][]{Todt2015,wrh2004,Hamann1998}. The PoWR code solves the
non-LTE radiative transfer in a spherically expanding atmosphere
simultaneously with the statistical equilibrium equations and accounts
at the same time for energy conservation. Complex model atoms with
hundreds of levels and thousands of transitions are taken into account.
Iron and iron-group elements with millions of lines are included
through the concept of super-levels \citep{graf2002}.  An  X-ray field
with the observed intensity is artificially added to account for its
ionizing effect \citep{Baum1992}.  Radiation  pressure is consistently
included in the treatment of the photosphere, hence providing a
realistic description of the photosphere-wind transition region
\citep{Sander2015}.

\begin{figure}[t]
  \begin{center}
\includegraphics[width=1.0\columnwidth]{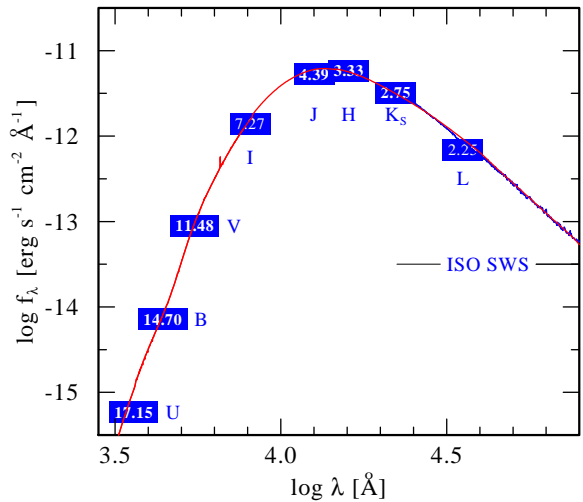}
\caption{
Spectral energy distribution (SED) for \cyg; photometric measurements 
in UBV and JHK bands are represented by blue boxes with the magnitudes 
imprinted.   The synthetic SED from a PoWR model with parameters from 
Table \ref{tab:par} is shown as a red solid line. 
}
\label{fig:sed}
\end{center}
\end{figure}

With the parameters and abundances (enhanced nitrogen, depleted carbon and 
oxygen) adopted from \citet{Clark2012}, 
the synthetic spectrum obtained with PoWR compares well with the observed 
optical spectrum of \cyg\ published by the same authors and by
\cite{Maiza2016}, thus confirming their analysis. Our model H$\alpha$
line is also in good  agreement with the observation shown in figure\,1
by  \citet{Clark2012}\footnote{Monitoring observations show significant
H$\alpha$ variability \citep[][see Sect.\,\ref{sec:sv}]{Chentsov2013}}. 
Our fit of H$\alpha$ and the other Balmer lines required a
clumping factor $D=6$ in the wind. \citet{Clark2012} mention that they
applied a clumping value (defined as the inverse of our
clumping factor $D$) of 0.04 at final velocity, but with a radial onset 
of clumping at 200\,km\,s$^{-1}$. This seems to result in a similar
degree of clumping as our model in those regions where Balmer emissions form.
One must keep in mind that there is a degeneracy between
clumping factor (at relevant layers) and mass-loss rate when fitting
recombination-fed emission lines \citep{Hamann1998}.

In a recent work, 
\citet{Whittet2015} reassessed the interstellar environments and
dust properties toward \cyg\ and obtained $R_V=3.05 \pm 0.1$ and
$E_{{\rm B}-{\rm V}} = 3.33$\,mag. 
Comparing the photometric measurements with the spectral  energy
distribution from the model (see Fig.\,\ref{fig:sed}), we obtained good
agreement when adopting these values\footnote{\citet{Clark2012} 
used $E_{{\rm B}-{\rm V}} =
3.84$\,mag with a reddening parameter  $R_\mathrm{V} = 2.65$}.  

Our slightly lower reddening leads to a  bolometric luminosity of 
$\log{L_\mathrm{bol}/L_\odot}=6.22$ compared to 6.28 by
\citet{Clark2012}. The set of fundamental parameters that we adopt in
the following is compiled in Table\,\ref{tab:par}.

With the help of the PoWR model we investigated  whether the wind of
\cyg\ can be radiatively driven. The PoWR models compute the work ratio
$Q$ defined as the mechanical work per unit time done by the radiation
field as compared to the mechanical luminosity of the wind
\begin{equation}
  Q\equiv \frac{\displaystyle{\int_r} \left[g_\mathrm{ rad}(r) - \frac{1}{\rho(r)}
      \frac{\mathrm{ d}P_\mathrm{ g}}{\mathrm{ d}r}\right]\mathrm{ d}r}
  {\displaystyle{\int_r} \left[v\frac{\mathrm{ d}v}{\mathrm{ d}r} +
      \frac{GM_\ast}{r^2}\right]\mathrm{ d}r},
  \label{eq:q}
\end{equation}
where $g_\mathrm{rad}$ is the radiative acceleration, $P_\mathrm{g}$
is the gas pressure, and other symbols have their usual meanings. A
hydrodynamically consistent model must give $Q=1$. When $Q>1$ the
model predicts that the radiation pressure should actually drive a 
stronger wind (i.e.\ with higher $\dot{M}$ and/or $v_\infty$). 
Correspondingly, when $Q<1$ the model indicates that the  radiative
acceleration is not sufficient for driving a wind with  the adopted
parameters.

\begin{figure}[t!]
  \begin{center}
    \includegraphics[width=0.99\columnwidth]{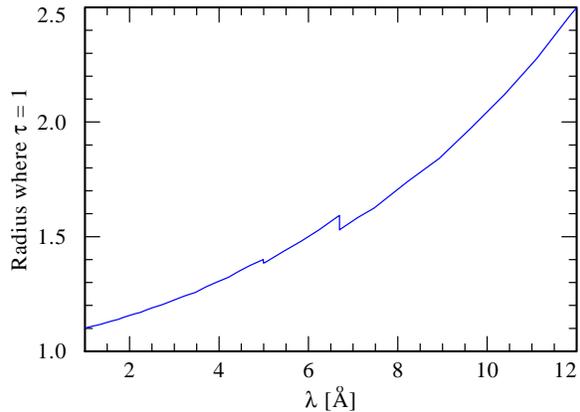}
    \caption{Radius (in units of $R_\ast$) where the continuum optical 
    depth reaches unity, as function of wavelength in the X-ray
    range. Calculated from the PoWR model for 
     the ``cool'' wind component of \cyg\ with the parameters from
     Table\,\ref{tab:par}}
    \label{fig:t1}
  \end{center}
\end{figure}

Using the model parameters from Table\,\ref{tab:par}, we computed this 
work ratio and obtained $Q\approx 1$, implying that the  stellar wind
of \cyg\ wind is consistent with being  radiatively driven. As a test,
we computed also models with $v_\infty=1000\kms$. The effect of a higher
terminal speed for the lines in the optical part of the spectrum is
marginal; however, the work ratio Q becomes significantly smaller than
unity, implying that such a high wind velocity could not be 
maintained by radiative driving. 

Hence we adopt $v_\infty = 400\kms$ as most consistent. Taking this as 
an upper limit for possible velocity discontinuities,
the strong-shock condition yields $6\mk$ for the maximum
temperature a shock could produce.

The main goal of the present study is to analyze the X-ray spectrum of
\cyg. For this task, a detailed knowledge of stellar wind opacities and
the radiation field is required.  To compute these quantities, we
employed the   PoWR code to compute the ``cool'' wind opacity. 
{\changed This is sufficient since no signatures of absorption in the 
``hot'' X-ray emitting plasma  is seen in X-ray spectra, the hot plasma 
component is, thus, optically thin.} 
Figure\,\ref{fig:t1} shows the radius in the wind
where the optical depth for X-rays becomes unity.  For the X-rays at
wavelengths longward of $\approx 8$\,\AA, the wind is optically thick
below $\approx 2\,R_\ast$. Therefore, if X-ray emission were produced
below this radius, one would expect severe wind absorption
\citep{Cas1981,Ignace2000}.  

\subsection{Spectral and photometric variability at optical and radio 
wavelengths}
\label{sec:sv}

\begin{figure*}[ht!]
\begin{center}
\includegraphics[width=0.98\textwidth]{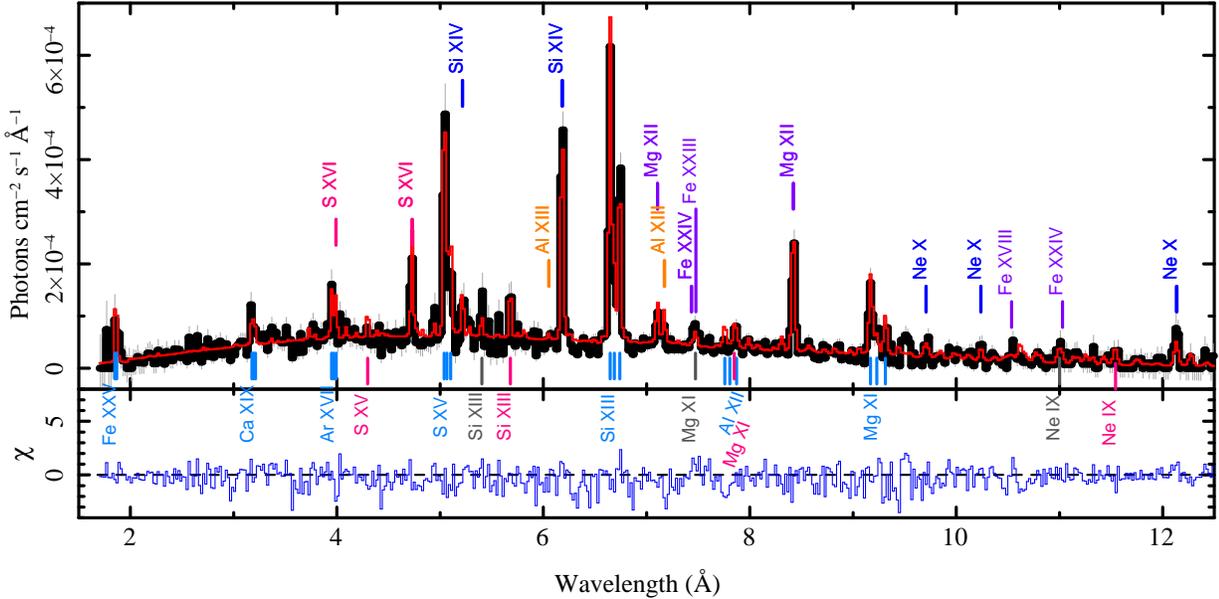}
\caption{The \chan\ HETGS spectrum of \cyg\ with prominent lines 
identified. Lines from He-like ions are marked below the spectrum, while
H-like and other Fe lines are marked above the spectrum.  The
observation is in black, with a model in red, and residuals in the lower
panel in blue.  For illustration only, counts were converted to photon flux
using the instrument responses (but still include the instrumental
resolution), combining the positive and negative first orders of
the spectrum from the HEG and MEG arm. 
}
\label{fig:hetgspec}
\end{center}
\end{figure*}

\cyg\ is a well-known variable. \citet{Gottlieb1978} noticed  an
irregular variability with by $\Delta B\approx 0.3$\,mag. Among other
targets, \citet{Laur2012} also observed \cyg\ for 300 days in 2011 and
confirmed its irregular variability. No clear period was detected, but
a time scale for variability of the order of 30 days was established. 
They found a mild trend in the observed $V-I$ color, and suggested that
this could be a manifestation of spectral-type variability.
\citet{Morford2016} reported \cyg\ variability at radio wavelengths.  


\begin{table*}[t]
  \begin{center}
    \caption{Parameters from simultaneous fits
      of the MEG, HEG, and zeroth-order spectra of \cyg\ in the
      $0.6$--$7.0\kev$ range with three- or two-temperature {\em apec}
      plasma models accounting for interstellar absorption (via {\it phabs})
      \citep{BCMC:1992,Smith2001}. 
} \label{tbl:specfit}
\small
\begin{tabular}[t]{l r@{\,$\pm$}l r@{\,$\pm$}l }  
\hline \hline 
\rule[-1mm]{0mm}{4.2mm} & \multicolumn{2}{c}{Three-temp.} 
                        & \multicolumn{2}{c}{Two-temp.} \\
\hline 
$N_{\rm H}\,[10^{22}\mathrm{cm^{-2}}]$ \rule[-1mm]{0mm}{4.5mm}
                        & 2.05 & 0.05$^{({\rm a})}$ & 1.8 & 0.03 \\[1mm]
$kT_1\,[\mathrm{keV}]$  & 0.20 & 0.04  & \multicolumn{2}{c}{---}  \\
$kT_2\,[\mathrm{keV}]$  & 0.81 & 0.1   & 0.77 & 0.1 \\
$kT_3\,[\mathrm{keV}]$  & 1.86 & 0.2   & 1.95 & 0.2 \\[1mm]
$EM_1\,[10^{56}\cmmthree]$& 33.89 &13.60 & \multicolumn{2}{c}{---} \\
$EM_2\,[10^{56}\cmmthree]$&  2.50 & 0.34 & 2.60 & 0.34 \\
$EM_3\,[10^{56}\cmmthree]$&  1.01 & 0.30 & 1.10 & 0.30 \\[1mm]

$f_{\rm x}\,[10^{-12}\mathrm{erg\,cm^{-2}s^{-1}}$]&
   \multicolumn{4}{c}{1.9\,$\pm$0.2$^{({\rm b})}$}\\[1mm]
        %
        %
        \hline
\multicolumn{5}{l}{$^{\rm (a)}$Error margins refer to $1\sigma$
uncertainty\rule[-1mm]{0mm}{5mm}}\\
\multicolumn{5}{l}{$^{\rm (b)}$ Observed flux in the $0.6$--$7.0\kev$ band}\\
        %
      \end{tabular}
  \end{center}
\end{table*}

\citet{Salas2015} conducted a 1.5-year-long photometric study of
variability of stars in the Cygnus~OB2 association. They concluded
that \cyg\ is an irregular or long-period variable with a period of 
54~days, which is a factor of 10 longer than our estimate of the wind
flow-time. The light curve of \cyg\ in the $I$-band exhibits changes
with an amplitude of $\Delta\mathrm{I}=0.18\,\mathrm{mag}$.

Besides photometric variability, \cyg\ also shows spectral variability.
\citet{Souza1980} found evidence for spectral and radial-velocity
changes. In particular, the H$\alpha$ line centroid moves by more than
$30\kms$. \citet{Klochkova2004} and \citet{Chentsov2013} presented a
time series of high-resolution spectra of \cyg. They found
spectroscopic manifestations of an unstable stellar wind, namely line
profile asymmetries and variations that differ  from line to line. 
They suggested that the H$\alpha$ line profile indicates that some
fraction of the wind falls back onto the star.

\citet{Clark2012} gave a detailed review of \cyg's variability, and
concluded that there is no significant evidence for a long-term
evolution of the spectral type over the past 50~years.  Short term 
variability as observed in \cyg\ is commonly seen in other
luminous blue hyper- and supergiants as well.

{\changed \citet{Scuderi1998} measured radio spectrum of \cyg\ and 
concluded that it fully consistent with being thermal.  Recently, 
\citet{Morford2016} obtained the first ever resolved 
detection of \cyg\ at 21\,cm and measured unclumped mass-loss rate 
$\dot{M}\approx 5.4\times 10^{-6}$.  Furthermore, they 
observed 50\,\%\, increase in the mass-loss rate of over the 14\,d period and 
discussed previous detections of radio flux variability.}


\subsection{Binarity status}
\label{sect:binary}

The multiplicity of \cyg\ was carefully investigated by many authors as
a possible explanation for its outstanding luminosity and variability.
From the analysis of spectroscopic time series, \citet{Klochkova2004}
and \citet{Chentsov2013} excluded \cyg\ as a double-lined spectroscopic
binary, but they could not principally rule out a possible binarity. 

\citet{Caballero2014} conducted a high angular resolution survey of
massive OB  stars in the Cygnus OB2 association using the fine guidance
sensor of the  {\em Hubble Space Telescope (HST)}. A highlight of
this study was the discovery of a companion to \cyg\ with a separation
of $63.6\,\mathrm{mas}$. Under the assumption that the projected
separation corresponds to the apastron separation, and for a total
system mass of $120\msun$, the orbital period would be $\approx
30\,\mathrm{yr}$ {\changed and the orbital separation {\changed $\approx 
104\,R_\ast$ 
or $\approx 110$\,au}.}
The secondary is $\Delta V \approx 2.3$ fainter than
the primary. 

\citet{Mar2016} used speckle interferometry and confirmed the detection
of a second component in \cyg. They were able to measure the changes in
the position angle of the secondary component, and suggested that the
binary period is $\sim 100$\,yr. They discovered of an even fainter 
third component in the \cyg\ system.


The brightness ratio ($\Delta V \approx 2.3$) suggests that the
secondary, most likely also a hot star, must have a much smaller radius
(e.g.\ like an OB star of lower luminosity class).  The wind of such
putative companion is expected to be quite fast. For the purpose of an
estimate, we might imagine an O9.5\,II star with parameters as
recently derived for  $\delta$\,Ori: $\log{L_\mathrm{bol}/L_\odot}
\approx 4.8$, $\dot{M}\approx 2\times 10^{-7}\,M_\odot$\,yr$^{-1}$, and
$v_\infty\approx 2000$\,km\,s$^{-1}$ \citep{Shenar2015}.   When such a
fast wind collides with the denser and slower  wind of a B hypergiant,
strong X-ray emission and high shock temperatures should result 
\citep[e.g.][]{Stevens1992}.


\section{\chan\ observations of \cyg} \label{sec:obs}

In this paper we report X-ray observations\footnote{\chan\ Observation
Identifier 16659} of \cyg\ obtained on 2015-01-14 with an exposure time
of $138\ks$ using the \cxo\ HETG spectrometer \citep{Canizares2005}. The
HETGS spectra cover a wavelength range from about 1 to 30\,\AA, as dispersed by
two types of grating facets, the High Energy Grating (HEG) and the
Medium Energy Grating (MEG), with resolving powers ranging from 100 to
1000, and an approximately constant FWHM of 0.012\,\AA\ for HEG and
0.023\,\AA\ for MEG.

The \cxo\ data were reprocessed with standard \cxo\ Interactive
Analysis of Observations (CIAO) programs \citep{Fruscione2006} to
apply the most recent calibration data (CIAO version 4.6 and
calibration database version 4.6.5).  The data are thus
composed of four orders per source per observation: the positive and
negative first orders for each grating (MEG and HEG), which
have different efficiencies and resolving powers. The default binning
over-samples the instrumental resolution by about a factor of four.
The merged HETGS flux spectrum is shown in Fig.~\ref{fig:hetgspec}.

\begin{figure}[h]
  \centering\leavevmode
  \includegraphics*[width=1.0\columnwidth]
                   {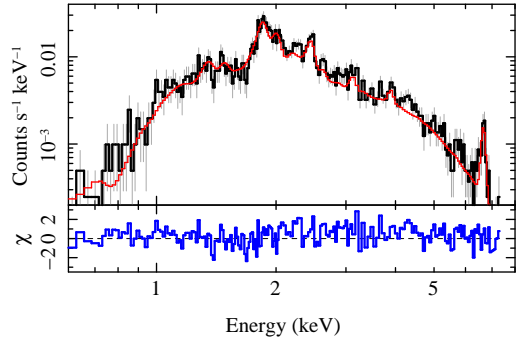}
    \caption{HETGS zeroth order count spectrum (black histogram with
    gray errorbars), plasma model (red),
    and residuals (lower panel, blue)}
   \label{fig:zo}
\end{figure}

When operating together with HETGS, the ACIS-S also simultaneously
obtains a zeroth-order, low-resolution X-ray spectrum which can be a useful
complement to the high-resolution spectrum, especially in the vicinity
of Fe~K, though care must be taken to assess photon event pileup.
Figure\,\ref{fig:zo} shows the zeroth-order spectrum of \cyg.

We modeled the spectrum primarily using the {\it Interactive Spectral
Interpretation System} \citep[][]{Houck:00} which implements
interfaces to the {\it AtomDB} \citep{Foster:Smith:Brickhouse:2012}
and to {\it XSPEC} models \citep{XSPEC1996}.  We fit a 3-temperature
{\em apec} model to the high-resolution and the zeroth-order spectrum.
We allowed relative abundances of prominent species (Mg, Si, S, Fe) to
float since this gave a somewhat better fit. Formally, the best fit
the the X-ray spectrum (as shown in the figures of this paper) was
obtained with slightly sub-solar abundances (factors compared to solar:
0.6 for Mg, 0.7 for Si, 0.9 for S, and 0.8 for Fe).
This could be in part due to degeneracies between discrete temperature
components, abundances, and absorption (both line-of-sight, and
wind-intrinsic), an not be considered to be real. 

Since the absorption is large, $N_\mathrm{H} \sim 10^{22}\cmmtwo$, the
lowest-temperature component in the tree-temperature model 
is poorly constrained. Using a
two-temperature model we obtain a fit of similar quality with a slightly lower
absorption and lower temperature for the middle component. This
illustrates some of the degeneracy in global modeling.  Either fits 
can serve as equally good basis for detailed line measurement by
providing a continuum model and approximate temperatures.  We list the
model parameters in Table~\ref{tbl:specfit}.  The models are in  good
agreement with those previously published based on the analysis of
low-spectral resolution \xmm\ spectra. The absorbing column derived
from spectral fitting corresponds well with the interstellar reddening, 
using the conversion factor $N_\mathrm{H} \approx E_{{\rm B}-{\rm
V}}\times  5.8~10^{21}$\,cm$^{-2}$ \citep{Bohlin1978}.

The observed flux in the $0.2$ -- $10\kev$ band is
$f_\mathrm{x}\approx1.9 \times 10^{-12}\eflux$. Using the column density
from the two-temperature fit, the (unabsorbed) model
luminosity becomes $L_\mathrm{x}\approx 6.8\times 10^{33}\lum$, or
$\log{L_\mathrm{x}/L_\mathrm{bol}} \approx -5.7$, which is factor of 
two larger than found in previous studies \citep[e.g.][]{Rauw2011}.

With the caveat that the contribution of the soft plasma components
might be underestimated, the emission measure weighted temperature of
the X-ray emitting plasma is $\approx 13\mk$
($1.1\kev$), while the hottest plasma component has a temperature of  
$\approx 22\mk$ ($1.9\kev$). This is significantly higher than could be 
explained by intrinsic shocks in the relatively slow wind of \cyg.  The high
temperature could be explained, however, by a collision of a fast wind
with $v\sim 1000\kms$ from a presumable OB-type companion with the
slow wind of the blue hypergiant. The simple colliding wind model
\citep[see Eqs.(1)--(4) in][]{Luehrs1997} allows us to crudely
estimate the possible location of the wind-wind collision region in
\cyg. {\changed The colliding wind zone is expected to be concave around the 
O-type component, with the apex of the colliding wind cone located at about
$\approx 40$\,au from the secondary  ($\approx 1000 R_{\ast,{\rm O}}$ assuming 
$R_{\ast,{\rm O}}=9\,R_\odot$) and $\approx 70$\,au ($\approx 66 R_{\ast, 
{\rm BI^+}}$) from the primary. This is a prediction that, in
principle, could be checked by investigating the high-resolution X-ray
spectra (see Sect.\,\ref{sec:fir}).

In such wide binary the plasma cooling in the colliding wind zone is adiabatic. 
The resulting X-ray luminosity scales inversely with the binary 
separation \citep{Stevens1992}. Therefore, for a binary on elliptic orbit, the 
modulation of X-ray flux on the time scale associated with the binary period is 
expected. In case of \cyg, one would expect to observe X-ray variability on 
times scales $>30$\,yr. This prediction can also be checked observationally. }

\subsection{Temporal variations of the X-ray flux}

\begin{figure}[t]
  \begin{center}
    \includegraphics*[width=0.95\columnwidth, viewport=45 35 495
      460]{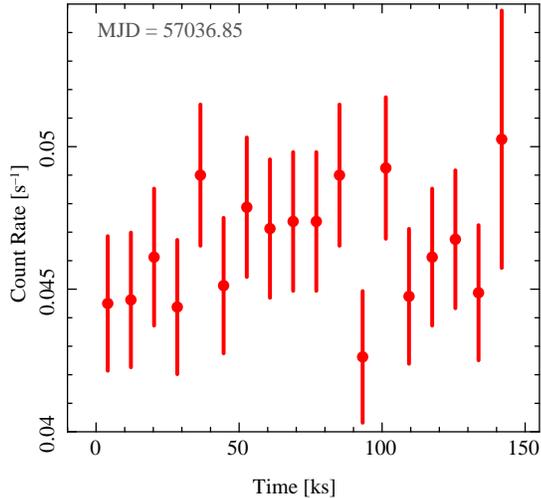}
\caption{The \cxo\ HETG X-ray light curve of \cyg\ in the $0.6$ --
$7.0\kev$ band during our observation on 2015-01-14. The data are
binned to 8\,ks. Error bars indicate 1\,$\sigma$ statistics. }
    \label{fig:lc1ks}
  \end{center}
\end{figure}

The X-ray light-curve during our \cxo\ HETG observation
($36\,\mathrm{h}$ exposure time) is shown in Fig.~\ref{fig:lc1ks}. It
is consistent with being constant, albeit the variance seems to
increase towards the end of the observation. Compared to the exposure
time, the characteristic wind flow time in \cyg\ is long,
$t_\mathrm{flow}=R_\ast v_\infty^{-1} \approx 120\,\mathrm{h}$ or
$5\,\mathrm{days}$. Any variability on a much shorter time scale thus
would be difficult to explain.

\cyg\ has been sporadically observed in X-rays since this range became
accessible \citep{Harnden1979, Kitamoto1996}.  \citet{Waldron1998}
pointed out that in the 1980s and 1990s, the X-ray emission of \cyg\
may have been steadily increasing at a slow rate.  They also reported
short-term variability at the level of 20\%, but were not able to
firmly attribute this variability to the star itself (due to suspected
instrumental effects).

\citet{AC2007} reported a roughly linear decrease of the \chan\ ACIS
count rate of \cyg\ from $\approx 0.18\,\mathrm{count\,s^{-1}}$ to
$\approx 0.16\,\mathrm{count\,s^{-1}}$ during a $98\ks$ exposure.

\citet{Rauw2011} analyzed six \xmm\ observations of the Cyg\ OB2
region (four in 2004 and two in 2007) with a total exposure time of
$148\ks$. \cyg\ revealed variability at a 10\%\ level with timescales
from a few days to a few weeks. A larger variation of the X-ray flux
(40\%) was seen between observations made in 2004 and 2007. These
variations were attributed to changes in the column density of
absorption, while the plasma temperature was found to be relatively
constant. The \cyg\ X-ray spectra were fitted best by 
multi-temperature plasma models with $kT_1=0.76 \pm 0.03$ and
$kT_2=2.03 \pm 0.19\kev$.

\begin{figure}[t!]
\begin{center}
\includegraphics[width=1.0\columnwidth]{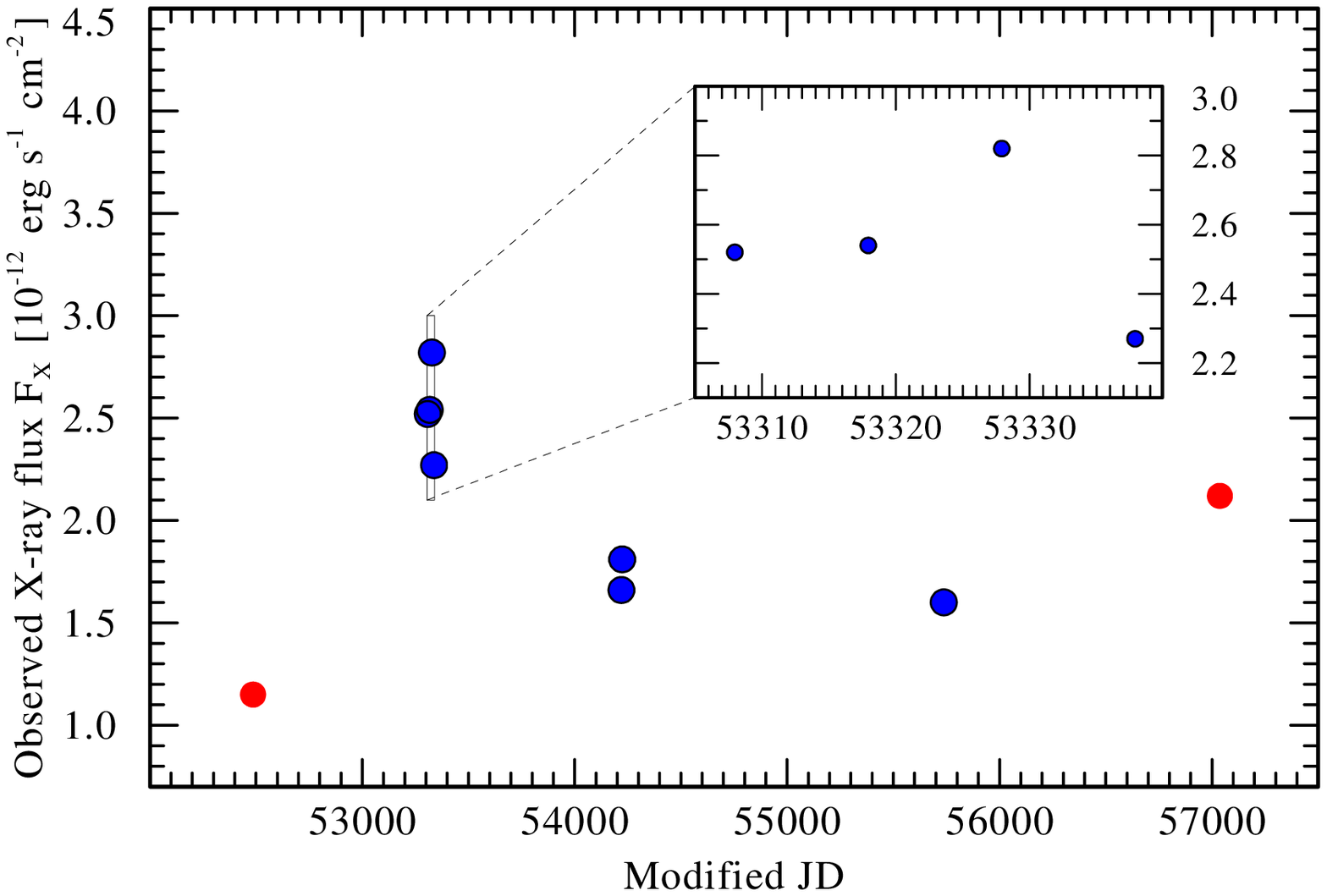}
\caption{X-ray flux of \cyg\ in the $0.5$--$10\kev$ band as measured
at different epochs within $\sim 10\,\mathrm{yr}$. All data points
except the first and the last show X-ray fluxes from \xmm\
observations, adopted from \citet{Cazorla2014}. The first red point is from 
a serendipitous \cxo\ observation, while the last red point represents our
\cxo\ observation on 2015-01-14 (MJD\,57036). The error bars are
smaller than the size of the symbols. The zoom demonstrates that
significant X-ray variability is also observed on a shorter time scale
($\approx 30$\,days).
    }
    \label{fig:longlc}
  \end{center}
\end{figure}

Further investigating the temporal evolution of the X-ray flux from
\cyg, \citet{Cazorla2014} also included observations obtained by the
{\em Swift} and {\em Suzaku} X-ray telescopes. A decrease of X-ray flux
between 2004 and 2011 {\changed (MJD\,53000-- 56000)} by 40\%\ earlier noticed 
by \citet{Rauw2011} was
confirmed.

Our latest observation yields a flux
that is 33\%\ higher than the previous \xmm\ observation obtained on
2011-06-25 {\changed (MJD\,55737)}, thus showing the opposite trend than 
before.  

We have additionally derived the flux from the earliest, serendipitous
\cxo\ HETG observation (Observation Identifier 2572, observed in 2002);
while far off-axis and not useful for high-resolution analysis, it is
sufficient to determine a flux of $1\times10^{-12}\eflux$ from the
nearly 1000 counts in the two dispersed spectra on the detector array.
Other previous \cxo\ observations of \cyg\ without grating were
affected by pile-up \citep{Rauw2015} and are therefore not useful for
flux estimates. 

Figure\,\ref{fig:longlc} shows the evolution of the X-ray flux from \cyg\
over time, combining our recent observation with all useful 
previous data from \cxo\ as well as from \xmm\footnote{the 
two instruments are cross-calibrated within $\sim 10$\%, see 
\url{xmm2.esac.esa.int/external/xmm_sw_cal/calib/documentation/index.shtml}}. 
Obviously, the X-ray flux varied by up to a factor of two. 

If the X-rays from \cyg\ were powered by wind-wind collision  in a
binary, modulations on the orbital time scale would be expected,
especially if the orbit was eccentric.  The recent discovery of a
binary companion with $\msim 30$\,yr period
(cf.\ Sect.\,{\ref{sect:binary}) supports this expectation. However,
Fig.\,\ref{fig:longlc} does not clearly suggest such a regular
behavior. On the contrary, short-term variability has been observed,
too (see insert in Fig.\,\ref{fig:longlc}). As obvious from the first
four \xmm\ observations, the X-ray flux can change by $\sim 20$\%\
within just one month \citep{Rauw2011}. Hence, one cannot exclude that
the observed light curve reflects only random variability.

\begin{figure}[!b]
  \centering\leavevmode
  \includegraphics*[width=0.90\columnwidth]{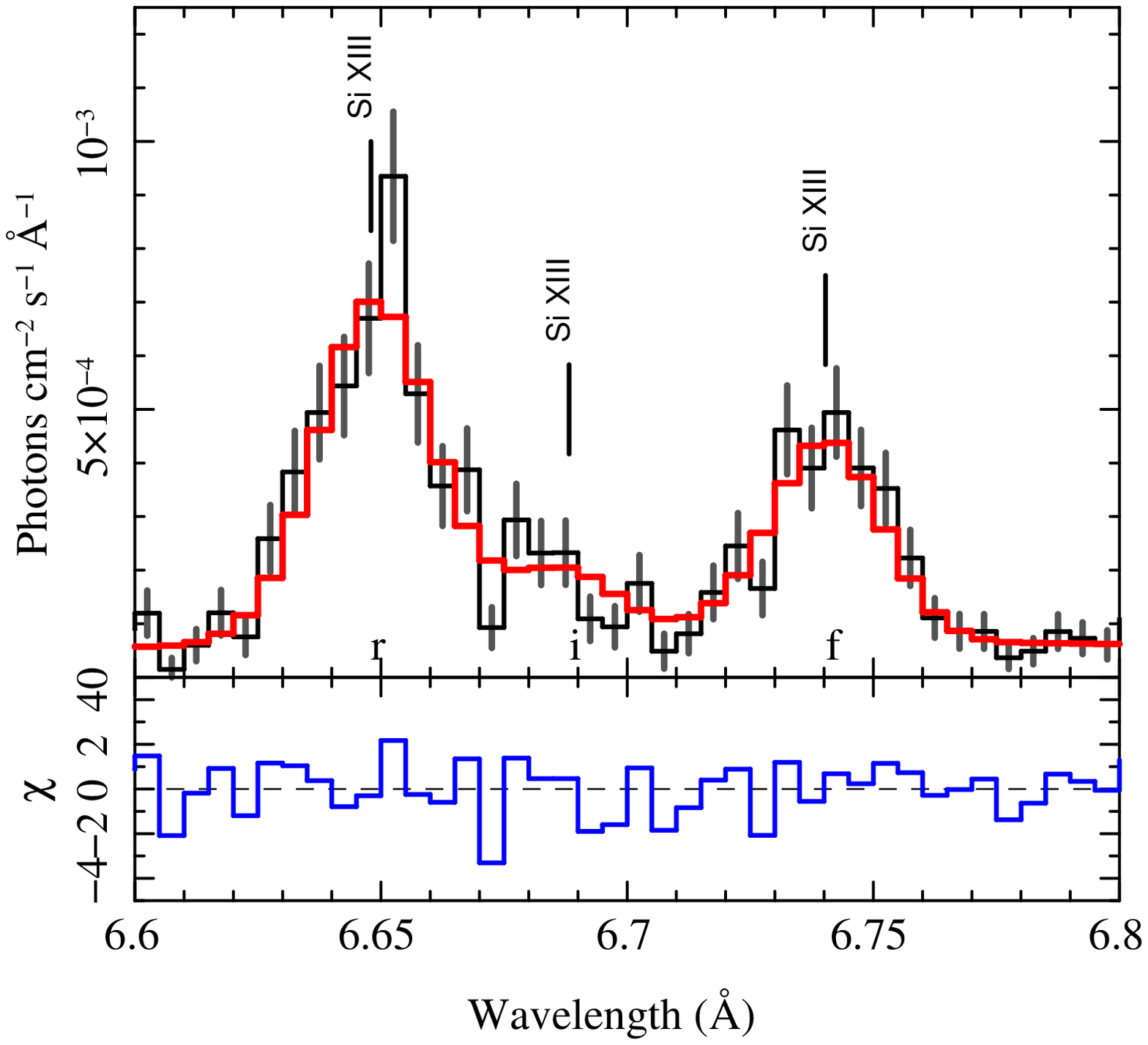}
  \vspace{3mm}\\
  \includegraphics*[width=0.90\columnwidth]{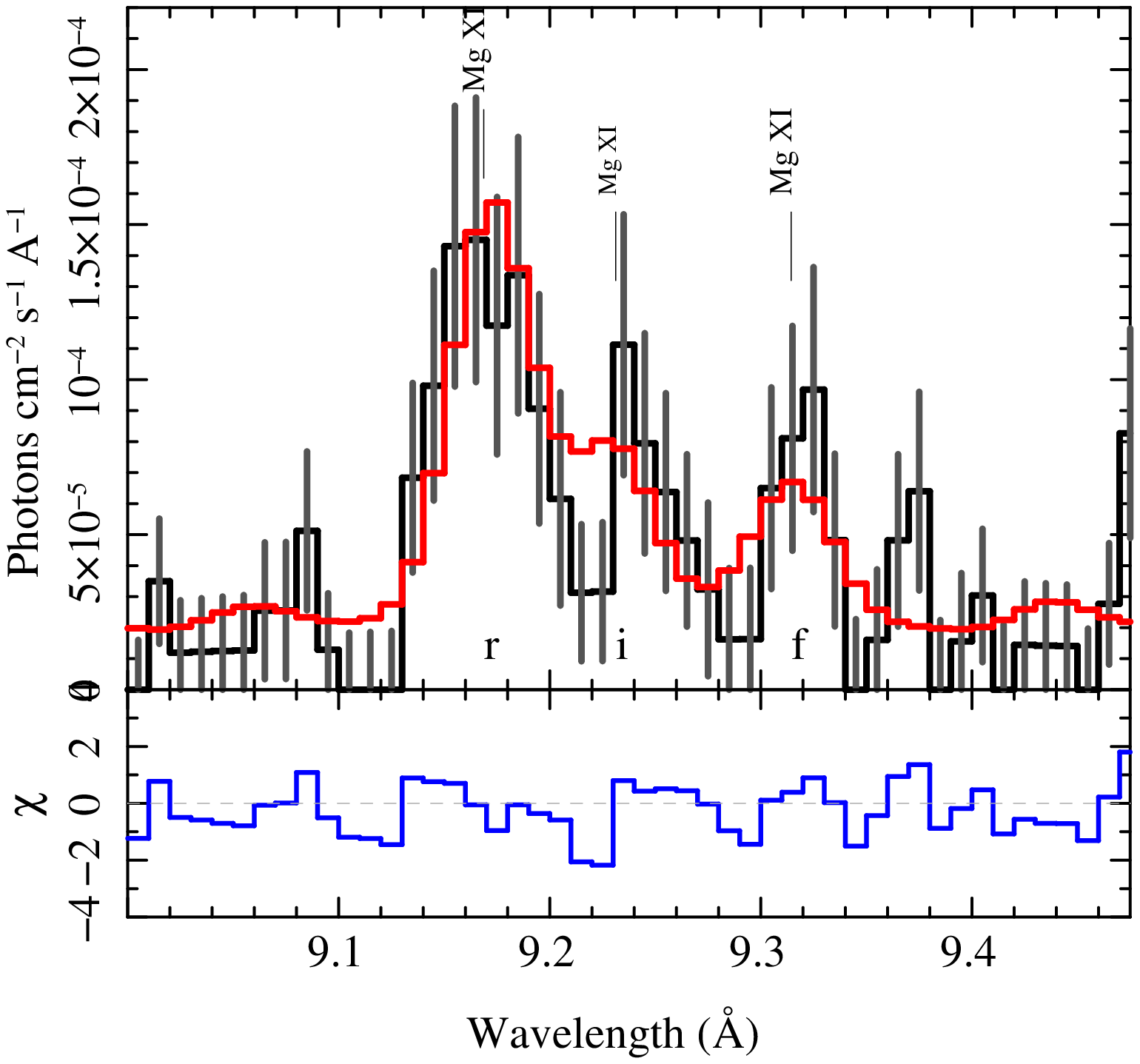}
\caption{The \eli{Si}{13} (top) and \eli{Mg}{11} (bottom) {\em fir}
triplets in the MEG first order spectra of \cyg.  In each plot, the top
panels show the photon count spectra as black histogram, with the best
fit model as the red histogram. The
rest-frame wavelengths the for resonance, intercombination, and forbidden
lines are marked.  Residuals are shown in the lower part of each panel.}
\label{fig:mgsihe}
\end{figure}

\subsection{Analysis of the X-ray emission line spectrum}
\label{sec:xanalysis}

\subsubsection{Lines of He-like ions}\label{sec:fir}

\label{sec:firanalysis}

The HETGS spectrum of \cyg\ is dominated by strong emission lines (see
Fig.\,\ref{fig:hetgspec}). Among them are the prominent lines of the
He-like ions \eli{Si}{13} and \eli{Mg}{11} (Fig.\,\ref{fig:mgsihe}). 
These ions show characteristic ``{\em fir} triplets'' of a forbidden
($z$), an intercombination ($x+y$), and a resonance ($w$) line
\citep{Gabriel1969}.

In order to measure the $f/i$ ratios, ${\cal R}(r)$, from this spectrum
as accurate as possible, we  used the global multi-temperature fit to
first provide an approximate plasma model. Then we fitted the lines
locally using the density-dependent emissivities for Mg or
Si.\footnote{See \url{http://space.mit.edu/cxc/analysis/he_modifier}
for details,  emissivity data, and code.} The fits are shown in
Fig.\,\ref{fig:mgsihe}, and the obtained values for ${\cal{R}}_\mathrm{obs}$ 
are given in Table\,\ref{tab:fir} together with their 90\% confidence
intervals.

Considering the theory, the ratio of fluxes between the forbidden and
intercombination lines, ${\cal{R}}(n_\mathrm{e}, T_\mathrm{rad}) =
z/(x+y)$, is sensitive to the UV radiative field and to the electron
density \citep[e.g.][]{porq2001}. A strong UV radiation leads to a
significant de-population of the upper level of the forbidden line to
the upper levels of the intercombination lines \citep{blum1972}.  For
the characteristic densities of OB and WR~star winds, this is the
dominant mechanism for forbidden line de-population
\citep[e.g.][]{WC2007,leu2007, osk2012}. Since the radiation field
dilutes with distance from the stellar surface, the ratio between
forbidden and intercombination line provides information about the
location of the X-ray emitting plasma.

\begin{table}[bth]
\begin{center}
\caption{Ratios ${\cal R}=f/i$ for He-like ions in the HETGS spectrum of \cyg}
\label{tab:fir}
    \vspace{1mm}
    \begin{tabular}[h]{lccc}  
\hline \hline 
      Ion & $\lambda(w)$\,[\AA]\rule[-1mm]{0mm}{5mm} & 
          ${\cal R_\mathrm{obs}}$ & $R_0$   \\
\hline 
      Si\,{\sc xiii}  & 6.65 &  3.22 (2.5\,...\,3.9) & 3.02 \\
      Mg\,{\sc xi}    & 9.17 &  0.78 (0.2\,...\,2.3) & 3.70  \\
\hline 
    \end{tabular}\\
  \end{center}
{\small{\bf Notes.} 
Wavelengths refer to the resonance lines ($w$ component).  For
${\cal{R}_{\rm obs}}$, the measured values are given together with
their 90\%\ confidence intervals intervals. The last column gives 
the asymptotic ratio $R_0$ that has been adopted for the calculation
(see text). 
}
\end{table}

The upper level of the forbidden line transition  can also be
de-populated by electron collisions. However, for being important 
this process requires electron densities that are comparable to the
``critical'' value $n_{\rm c}$, which is $4 \times 10^{13}\,{\rm cm}^{-3}$ and 
$6\times10^{12}\,{\rm cm}^{-3}$ for \eli{Si}{13} and \eli{Mg}{11},
respectively \citep{blum1972}.   


\begin{figure}[!ht]
  \begin{center}
\includegraphics[angle=-90,width=1.0\columnwidth]{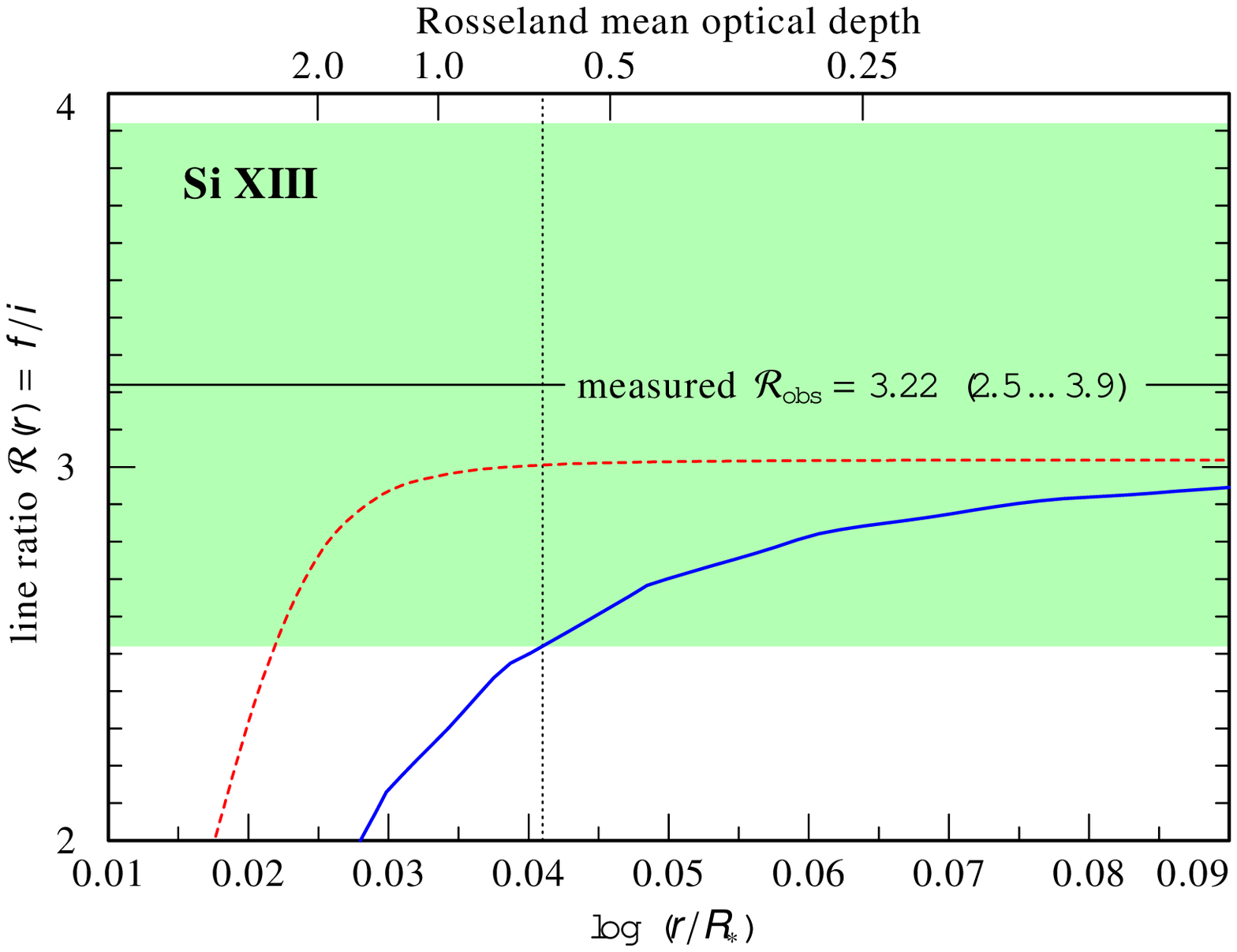}
\vspace{0mm}\\ 
\includegraphics[angle=-90,width=1.0\columnwidth]{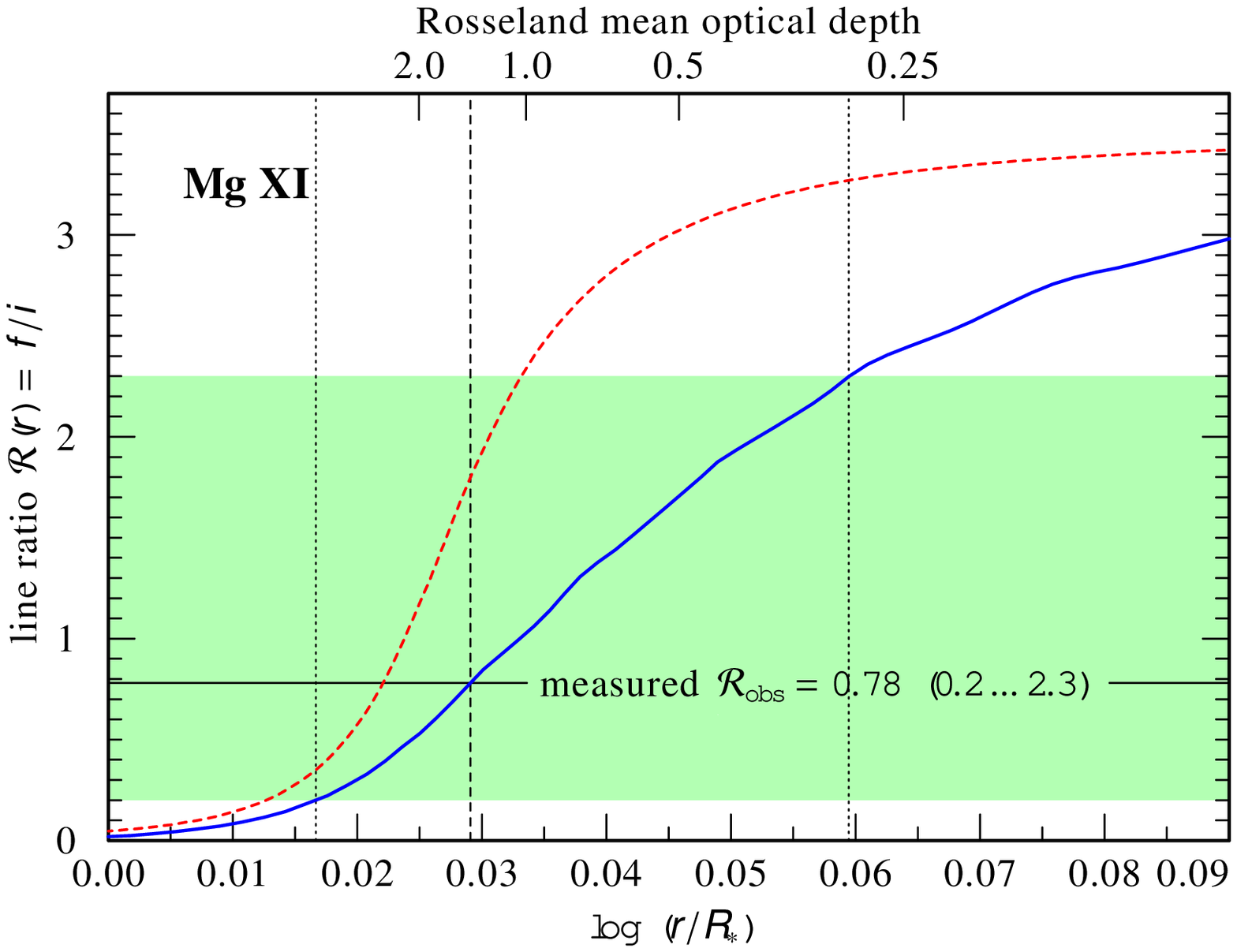}
\caption{
Theoretical ${\mathcal{R}}(r) = f/i$  as function of the radial location
of the X-ray emitting plasma for the {\em fir} triplet from \eli{Si}{13} 
(top panel) and \eli{Mg}{11} (bottom).  The red-dashed curves are 
computed assuming  that the de-population of the forbidden level is by
the photo-excitation only, while the blue solid curves includes the
contribution of collisions. In each panel, the measured value is
indicated by a horizontal  black line, while the green-shaded band
indicated its error margin. The intersection of ${\mathcal{R}}(r)$
(blue curve) with the  observed values ${{\mathcal{R}}}_{\mathrm{obs}}$
and their error margins are indicated by vertical dashed and dotted
lines, respectively.
}
\label{fig:simg}
\end{center}
\end{figure}

\cyg\ has relatively dense wind, low effective temperature,  and yet
a very hot X-ray plasma. For these conditions, it is not clear
{\it a priori}  which mechanism of forbidden line de-population
dominates.  Hence, in order to correctly apply the {\em fir}
diagnostic, we employed our PoWR model (cf.\ Sect.\,\ref{sec:windpar}).
For modeling  the $f/i$ ratio, we  follow the recipe by
\citet{Shenar2015} which is based on \citet{blum1972}.

%
%

For the wavelengths of the de-populating transitions (865\,\AA\ for
Si\,{\sc xiii}, 1034\,\AA\ for Mg\,{\sc xi}) we extract the mean 
intensities at each radial layer in the stellar wind, as provided by
our PoWR model calculation for \cyg. Note that all effects such as
diffuse emission, limb darkening, and attenuation of UV flux in the
wind are automatically  included in the model in a consistent manner.  
The density in the wind is also taken from our PoWR model. Taking the
same density for the collisional de-population relies on the
assumption that the densities in the shock-heated plasma is similar to
the smooth-wind density at the same radius. This is a reasonable 
first approximation, since hydrodynamical simulations 
for wind-embedded shocks powered by the line-driving instability did not
predict large over-densities in the shocked material \citep{feld1997}.

Furthermore, the modeling of the {\em f/i} ratio requires the relevant
atomic data. The transition  wavelengths and oscillator strengths are
extracted from the NIST database. The asymptotic value for the {\em
f/i} ratio, $R_0$,  which enters the the theoretical computation of
${\cal{R}}(r)$, is slightly temperature-dependent; the values adopted
here are read off from  figure\,8 in \citet{Porquet2000} for the
highest temperatures provided. 

The results of our {\em f/i} analysis are illustrated in 
Fig.~\ref{fig:simg}. For the He-like ions Si\,{\sc xiii} and Mg\,{\sc
xi}, the predicted {\em f/i} ratio ${\cal{R}}(r)$ is plotted as function
of the radial location $r$ of the emitting plasma. 

{\changed Different temperature plasma components (with 6\,MK and 20\,MK) may 
contribute significantly to the emission in Mg\,{\sc xi} and Mg\,{\sc xii} 
lines.  A better measurement of the S\,{\sc xv}, Si\,{\sc xiii},  and 
Mg\,{\sc xi} line ratios would help to constrain the relative contributions of 
plasma components to the line spectrum\footnote{these arguments are suggested 
by the reviewer}.}

First we test whether collisional or radiative excitation is the
dominant process for the forbidden line de-population. The calculations
reveal that in the {\em wind} of \cyg\  neither the UV field
(which is weak due to the low effective temperature of the star) nor
the electron collisions are able to de-populate the forbidden-line upper
state! With growing distance $r$, the predicted {\em f/i} ratio 
${\cal{R}}(r)$ soon approaches the limiting value $R_0$. Note that 
Fig.~\ref{fig:simg} zooms at the photosphere/wind transition region,
as can be recognized from the Rosseland-mean optical depth scale indicated on
the top of the diagrams\footnote{The stellar radius $R_\ast$
refers by our definition to $\tau_\mathrm{Rosseland} = 20$, while
$\tau_\mathrm{Rosseland} = 2/3$ is the radius from where most of the
photospheric flux escapes.}.   
The red-dotted line is calculated for zero
electron density, i.e.\ neglecting collisional de-population, while the
blue curves take collisional de-population into account.  The large
difference between these two curves reveals the leading role of 
collisions for this process in these layers. 

The (blue) theoretical curve may now be compared to the measured {\em
f/i} ratio ${{\mathcal{R}}}_{\mathrm{obs}}$, which is also indicated in
each of the panels of Fig.\,\ref{fig:simg} by a horizontal line 
together with the uncertainty of the measurements (90\% confidence
interval, green shaded band). 

For Si\,{\sc xiii} the measured ${\cal{R}}_\mathrm{obs}$ is consistent 
with the limiting $R_0$, i.e.\ with the absence of any de-populating
process. Hence, the X-ray emitting plasma could reside anywhere in the
wind, and its location cannot be constrained.  

A better constraint is provided by Mg\,{\sc xi}. As can be seen from
the lower panel of Fig.\,\ref{fig:simg}, the observed ratio of
forbidden to intercombination lines indicates that, at least within the
90\% confidence interval, the measured 
${{\mathcal{R}}}_{\mathrm{obs}}$ is {\em not} consistent with $R_0$,
i.e.\ de-population is required.  Densities that are comparable to
$n_\mathrm{c}$ are only encountered in or very close to the
photosphere. Hence we must conclude from Fig.\,\ref{fig:simg} that the
{\em fir} triplet of Mg\,{\sc xi} is emitted from plasma located at $r
<  1.15\,R_\ast$, i.e.\ practically directly at the photosphere.  
{\changed In this case, the X-ray emission lines shall be narrow and
would be unresolved even with HETGS grating spectrometry. This is 
because the wind expansion velocity at these radii is very small. Moreover, as 
can be seen in Fig.\,\ref{fig:t1}, below $1.8\,R_\ast$, stellar wind of \cyg\ 
is 
optically thick for radiation at $\lambda\,9.2$\,\AA.  Therefore, if 
radiation would originate in deeper wind layers, the spectral signatures of 
stellar wind absorption could be expected.  With all these in mind, the 
investigation of spectral lines should further help in constraining the hot 
plasma location.}

\begin{figure}[!tbh]
  \begin{center}
\includegraphics[width=1.0\columnwidth]{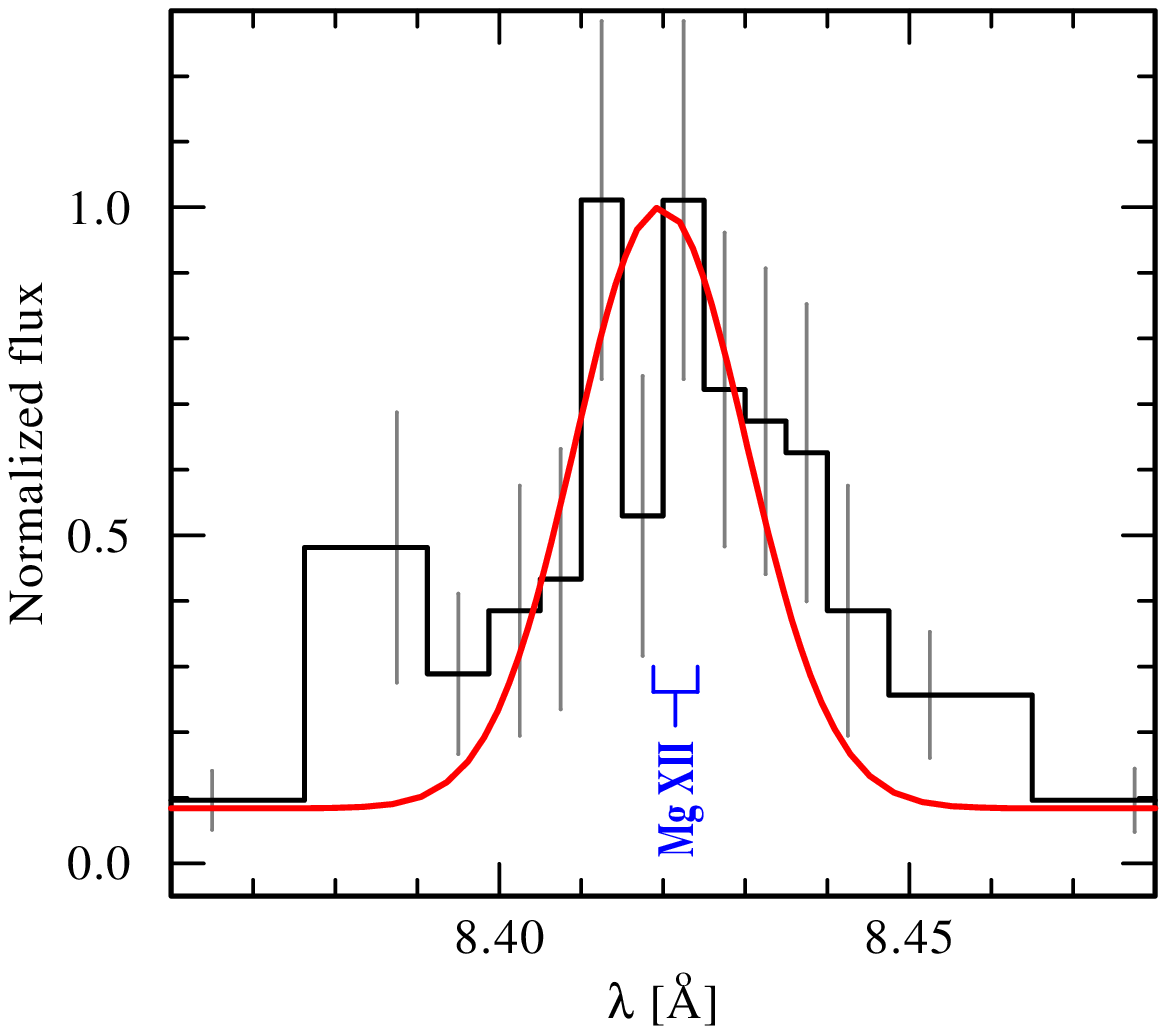}
\vspace{0mm}\\
\includegraphics[width=1.0\columnwidth]{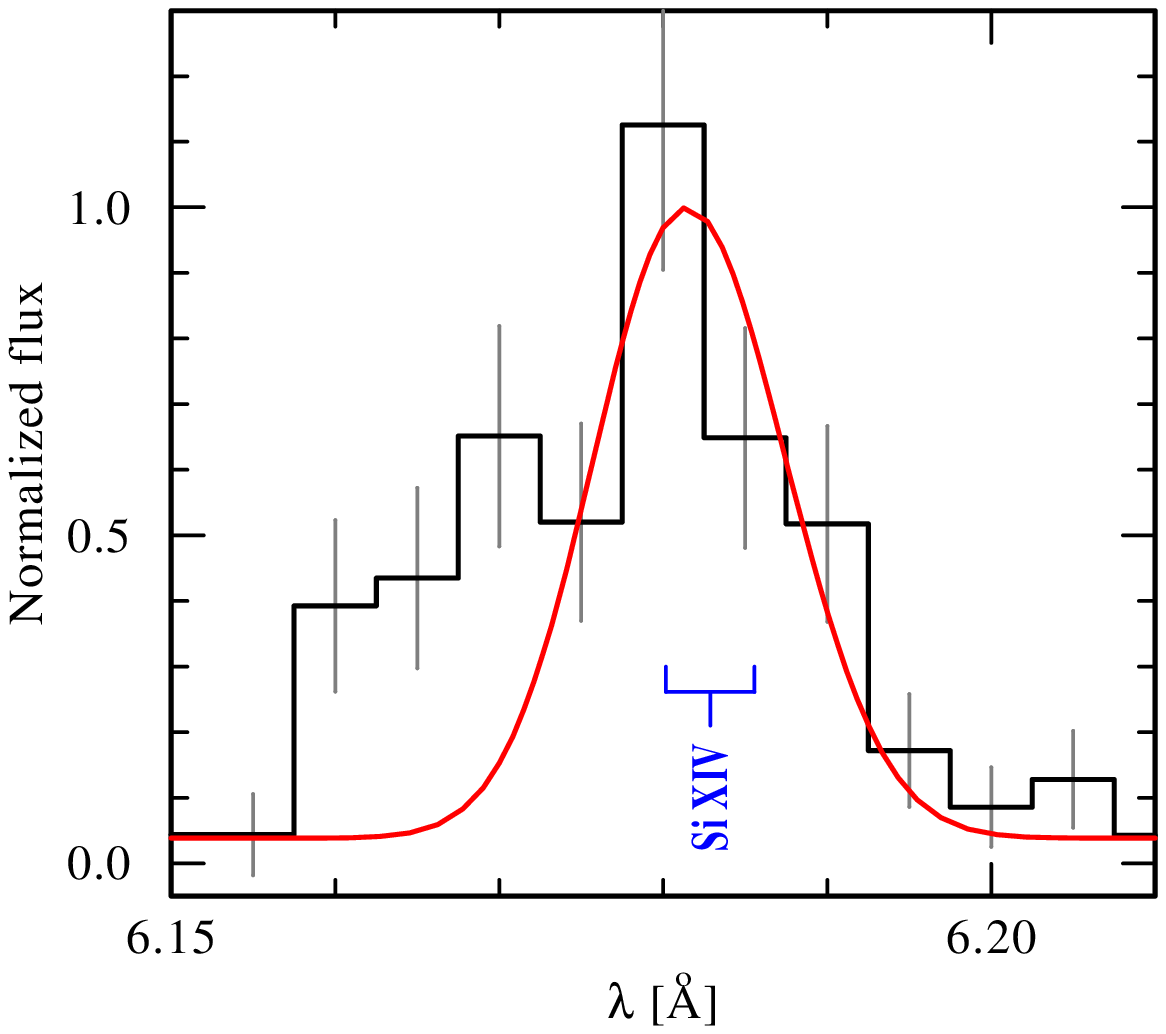}
\caption{Lines of Mg\,{\sc xii} (top panel) and Si\,{\sc xiv} (bottom) 
in the (co-added MEG$\pm  1$)
spectrum of \cyg\ (black histograms); the red curves shows a  model
profile assuming that the X-ray emitting plasma expands according to a
$\beta$-velocity law with $v_\infty = 400$\,km\,s$^-1$.  The rest frame
wavelengths of the doublet components are indicated in blue.}
\label{fig:mg12}
\end{center}
\end{figure}

\subsubsection{X-ray emission line profiles}
\label{sec:prof}

X-ray line profiles formed in a stellar wind are influenced by two
effects, the Doppler shift due to the wind expansion, and the 
absorption which is caused by the continuum opacity of the cool-wind
material.  The Doppler shifts can broaden the emission line profile up
to $\pm v_\infty$ \citep{macf1991,Ignace2001,Owocki:Cohen:2001}, while 
absorption and obscuration affect the back hemi\-sphere more than the
front hemisphere and thus cause the line profiles to become skewed and
effectively blue-shifted \citep{osk2006}. Especially if X-rays were
produced close to the photosphere, we would expect to find these
signatures of wind absorption. As can be seen from Fig.\,\ref{fig:t1},
at $1.18\,R_\ast$ the optical depth for X-rays exceeds already unity for any
wavelengths $>3$\,\AA\ and increases further with wavelength. At least 
such lines as Mg\,{\sc xii}\,$\lambda 8.42$\,\AA\ and Mg\,{\sc
xii}\,$\lambda 9.17$\,\AA, should show signs of wind absorption. 

To model the X-ray emission line profiles for \cyg, we perform the same
kind of calculations as described in \citet{osk2006}. The wind density 
and the opacities are again taken from our PoWR model for this star.
The main parameter of these calculations is the radius at which the
X-ray plasma appears; since the emissivity scales with the square of the
density, most of the photons are produced close to this onset radius. 

Based on the result from the {\em fir} analysis
(Sect.\,\ref{sec:firanalysis}), we adopt an onset radius for X-ray
emission close to the photosphere (its precise value is not relevant).  
The model then predicts that the profiles would not be broadened by the
wind velocity, because the latter is still tiny close to the
photosphere. The full width half maximum (FWHM) of the line profile
would be only $\approx 20$\,km\,s$^{-1}$, i.e.\ the observed profile
should only reflect the instrumental profile which is much broader. 


\begin{figure}[t]
  \begin{center}
    \includegraphics*[width=\columnwidth]{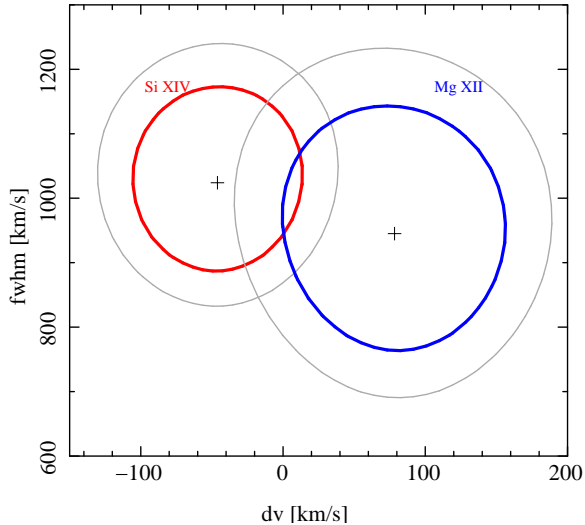}
\caption{Bold ovals show the $1\sigma$ confidence contours for the
Doppler-shift of the  centroid ($x$-axis) and and the width (FWHM,
$y$-axis) for the  \eli{Si}{14} (red) and the \eli{Mg}{12} (blue) line,
respectively. Gray contours refer to $90\%$
confidence. }
\label{fig:fw}
  \end{center}
\end{figure}

This expectation can be tested with the resonance lines of the
hydrogen-like ions Mg\,{\sc xii} at 8.42\,\AA\ and Si\,{\sc xiv} at
6.18\,\AA. Figure\,\ref{fig:mg12} shows that in both cases 
the observed profile is significantly broader than the
predicted profile convolved with the instrumental response. In other
words, the \chan\ HETGS observation resolves these profiles. 

To quantify the broadening of these lines, we fit their profiles with {\em 
apec} models over a small region of interest for each line, assuming that all 
lines in the region have the same Gaussian shape and a common Doppler shift.

The fits and confidence contours obtained with {\em the Interactive Spectral 
Interpretation System} software package 
are shown in  Fig.\,\ref{fig:fw}.  Both lines are 
intrinsically broadened. For both lines, the best fit indicates 
full-width-half-maximum (FWHM) values in excess of 700\,km\,s$^{-1}$. The 
lines are hardly Doppler-shifted against their rest wavelengths; centroid 
shifts by more than ~150\,km\,s$^{-1}$ can be ruled out.  {\nchange Such broad 
and unshifted lines are not expected from \cyg\ wind model.}

\section{Discussion and Summary}
\label{sec:disc}

\subsection{A scenario on origin of X-ray emission from \cyg. }

The analysis of new high-resolution X-ray spectra of \cyg\ combined
with modeling of its stellar wind by non-LTE stellar atmosphere 
provided fresh insights on the nature of \cyg. 
{\nchange Considering all observational facts together, the X-ray 
observations of \cyg\ are best explained by a colliding winds scenario.}

\medskip
\noindent
(1) The X-ray spectrum from \cyg\ is well described by a thermal plasma
in collisional equilibrium. Because of high inter-stellar
absorption, only the rather hard part of the spectrum is
observable. The highest temperature plasma components with 
$T_{\rm X} \msim 20$\,MK cannot be explained by 
intrinsic shocks in the relatively slow wind of \cyg\ ($v_\infty\approx
400\kms$).

\medskip
\noindent
(2) The X-ray luminosity is higher than expected from a
  mid-B spectral type.

\medskip
\noindent  
(3) The X-ray emission lines are broad. The line width (up to
  $1000\kms$) indicates that the X-ray emitting plasma moves
  with higher velocities than possible in the slow, cool wind 
  of \cyg\ ($v_\infty\approx 400\kms$).

  \medskip
\noindent
(4) The stellar wind opacities computed with our non-LTE models are
  high. If X-rays would to emerge from the inner wind, we would expect to
  see strong signs of wind attenuation in X-ray spectra. These are not 
observed.

  \medskip
\noindent
(5)  The non-LTE stellar atmosphere models indicate that collisions
  dominate over the UV de-excitation of the forbidden lines in 
   the He-like triplets seen in 
  \cyg's X-ray spectrum.  Hence the measured forbidden to
  intercombination line flux ratios provide information about plasma
  densities. Based on the analysis of {\changed Mg\,{\sc xi}}
  lines, we conclude that the X-ray plasma has the densities in excess of
  $10^{13}$\,cm$^{-3}$.  Such high densities could be achieved either
  at \cyg's photosphere, or in a hypothetical colliding wind region.

  {\changed 
(6) 
Assuming that the secondary is an O-type star, we estimate that the apex 
of the colliding wind shock is located at 
$\sim 1000\,R_\ast, {\rm O}$ from the secondary (see Section\,\ref{sec:obs}). 
At these distances, the UV flux from the secondary is strongly diluted, and 
does not dominate the forbidden line de-population. To fully understand X-ray 
emission from \cyg\
we need to better establish its binary properties.}

On the basis of the above points, and taking into account the recent
detection of a close-by companion to \cyg, the colliding wind
scenario seems to provide the most plausible explanation.  
{\changed This conclusion requires further quantitative testing
that will become possible only  when the secondary type and 
orbital parameters are known. Until then, the question whether the density
required for forbidden line depopulation can indeed be produced in the 
colliding wind zone of \cyg\ remains open. }

The high temperature plasma must occur in quite dense regions.  For 
a colliding wind zone, the pre-shock densities of 
the primary and the secondary winds are several orders of magnitude 
lower than needed to explain the $f/i$ line ratios. In 
radiative shock, high densities can be achieved in the post-shock cooling
zone, but those regions are deficient in X-ray emitting gas. 
Radiative shocks are known to be unstable via thin-shell instability
\citep{vishniac94}; perhaps some form of dynamical mixing allows
for both high densities and high temperatures to co-exist. 

{\changed It is informative to compare the $f/i$ ratios measured in \cyg\ (see 
Table\,\ref{tab:fir}) with those determined for the well known colliding wind 
binaries. E.g.\, the $f/i \approx 5$ ratio measured 
in the HETGS spectrum of Mg\,{\sc xi}  in the WR-binary WR\,140 is 
consistent with the absence of the forbidden line depopulation mechanism 
\citep{Pollock2005}. Similarly, in case of the LBV-binary 
$\eta$\,Car,  \citet{Henley2008} measured in the HETGS spectrum  of
Si\,{\sc xiii} the ratio $f/i>5$.} 

The hydrodynamic simulations of the massive binary 
$\eta$~Car display quite hot gas components and high densities
\citep{parkin2009}. $\eta$~Car involves a primary with a dense and 
relatively slow wind orbited by a secondary with a considerably faster 
wind. The models indicate that orbital motion
of the stars in $\eta$~Car helps to stabilize the wind collision
zone against thin-shell instabilities. The orbital period for 
$\eta$~Car is only $\sim 5$ years, the longer orbital period of \cyg\ 
could allow for stronger thin-shell instabilities and greater level 
of mixing resulting in the presence of hot, dense zones.

\subsection{On the binary evolutionary history of blue hypergiants}

The binary hypothesis is further supported by another
consideration.  Using archival X-ray data, we conducted a survey
of all 16 known Galactic blue hypergiants. We found that the majority
of these objects were not detected in X-rays. 
Some of the hypergiants, namely HD\,80077, Wd\,1-5, Wd\,1-13, and 
HD\,160529, were observed with modern telescopes and deep exposures, 
putting low upper limits on their
respective X-ray luminosities. For example, the upper limit for HD\,160529 
(B8-A9\,Ia$^+$)
 is $\log(L_\mathrm{x}/L_\mathrm{bol})<-8.5$
\citep{naze2012}. Especially interesting are the early-type
hypergiants HD\,169454 (B1\,Ia$^+$) and $\zeta^1$\,Sco (B1.5\,Ia$^+$),
both of which were not detected during the {\it ROSAT} All-Sky
Survey, with an upper limit for $\zeta^1\,$Sco at
$\log(L_\mathrm{x}/L_\mathrm{bol})<-8$. It appears safe to conclude
that blue hypergiants, in general, are not significant X-ray sources, with  
the X-ray luminosity not exceeding $\sim 10^{-9}$ of their bolometric 
luminosity. 

The two blue hypergiants which are outstandingly X-ray bright are  
BP\,Cru (B1\,Ia$^{+}$) and the subject of our study, \cyg\ (B3--4
Ia$^+$). Both of them are binaries. BP\,Cru is a well-known 
high-mass X-ray binary where a neutron star accretes the wind of 
its hypergiant companion \citep[e.g.][]{Kaper2006}. Its X-ray luminosity, 
$L_\mathrm{x}\sim 10^{37}\lum$ is determined by the accretion rate onto 
the neutron star. The X-ray luminosity of \cyg, however, is too low to 
suggest such a scenario. Except these two, no other binary hypergiants are 
confirmed so far. Albeit the numbers of known Galactic hypergiants is small, 
the very low binary fraction among them is in  stark contrast with the 
general OB star population. 

We suggest that all blue hypergiants are the products of binary evolution. 
\citet{Clark2014} discussed in detail the role of binarity in explaining the  
apparently single blue hypergiant Wd\,1-5. In case \cyg, the single star 
evolutionary models do not predict such extreme stars  
\citep{Ekstr2012}. Therefore, we propose that \cyg\ is either a former mass 
gainer in a very massive system with large initial mass ratio,  or a merger 
product. 

In the former case {\bf if} it would have been 
possible for the system to increase its orbital separation following the mass 
exchange, then presently fainter companion of \cyg\ may be the 
stripped remnant of the primary, e.g.\ a helium star of WR-type. Such star 
would have a strong and fast stellar wind, and help to explain the observed 
X-ray emission. However,  such a WR companion would have a spectrum dominated 
by emission lines. We carefully considered the spectrum of \cyg\ for such 
contamination, but we do not find any traces of WR features.    

It seems more plausible that \cyg\ is a result of a merger. Blue supergiant 
stars are expected to be mergers resulting from binary 
evolution \citep{Podsi1992}. We are not aware of detailed binary evolution 
models for blue hypergiants.  However, the low binary fraction among this type 
of stars along with their enhanced nitrogen  abundances \citep{Clark2012} may 
be a smoking gun pointing to their origin. 

We propose the following scenario. The progenitor of \cyg\ was initially 
a hierarchical multiple system, as common among massive stars 
\citep[e.g.][]{Shenar2015}. The eccentric Kozai-Lidov mechanism 
\citep{Naoz2014} caused  the strong  inclination and eccentricity fluctuations, 
resulting in tidal tightening of the inner binary. This inner tight binary 
merged and is observed today as the blue hypergiant. The initial tertiary is 
the present day binary component of \cyg. These two stars form a colliding wind 
system where the slow and dense wind of blue hypergiant collides with the fast 
wind of its late O-type companion. 

Furthermore, we showed that the single blue hypergiants are X-ray dim. If 
observed, bright X-ray emission from a blue hypergiant is a strong indicator 
of its binarity. We conclude that the majority of Galactic blue hypergiants are 
currently single stars but with previous binary evolutionary history.

\acknowledgments
{\changed We are grateful to the anonymous reviewer for her/his constructive 
comments that helped to improve the manuscript.}  This work has extensively 
used NASA Astrophysics Data System, and the SIMBAD database, operated at 
CDS, Strasbourg, France. This publication made use of data products provided
by the NASA/IPAC Infrared Science Archive. We are grateful to Dr.\,J. 
Ma{\'{\i}}z Apell{\'a}niz  for kindly providing us the optical spectrum of 
\cyg\ and to Dr. M. Mapelli for useful discussion on dynamics of multiple 
systems. LMO acknowledges support by DLR grant 50\,OR\,1302. AS is supported by 
the Deutsche Forschungsgemeinschaft (DFG) under grant HA 1455/26. DPH was 
supported by NASA grant GO5-16009A.}

{\it Facilities:}  \facility{CXO (HETG/ACIS)}.


\end{document}